\documentclass[journal,11pt]{IEEEtran}
\usepackage{graphicx}
\usepackage{float,lscape}
\usepackage{epstopdf}
\usepackage[hyphens]{url}
\usepackage{url}
\usepackage{algorithmic}
\usepackage{xr}
\usepackage{booktabs}
\usepackage{longtable}
\usepackage{tabularx}

\usepackage{multirow}
\usepackage{balance}
\usepackage{algorithmic}
\usepackage[table,xcdraw]{xcolor}
\usepackage[normalem]{ulem}
\useunder{\uline}{\ul}{}
\usepackage[]{algorithm2e}
\usepackage{ragged2e}
\usepackage{enumitem}
\usepackage{balance}
\usepackage{xcolor}
\usepackage{stfloats}
\ifCLASSOPTIONcompsoc
  \usepackage[nocompress]{cite}
\else
  \usepackage{cite}
\fi

\ifCLASSINFOpdf
\else
\fi

\begin{document}

\title{Artificial Intelligence and Machine Learning in 5G Network Security: Opportunities, advantages, and future research trends}

\author{Noman Haider$^{1}$, Zeeshan Baig$^{2}$, Muhammad Imran$^{3}$ \\
\begin{footnotesize}
$^{1}$College of Engineering and Science, Victoria University Sydney Campus, Sydney 2000, Australia.\\
$^{2}$Department of Computing, Faculty of Science and Engineering, Macquarie University, Sydney 2109, Australia.\\
$^{3}$College of Applied Computer Science, King Saud University, Riyadh, Saudi Arabia.\\
Corresponding author: noman90@ieee.org
\end{footnotesize}}

\markboth{Magazine type article}
{}

\IEEEtitleabstractindextext{
\begin{abstract}\justifying
Recent technological and architectural advancements in 5G networks have proven their worth as the deployment has started over the world. Key performance elevating factor from access to core network are softwareization, cloudification and virtualization of key enabling network functions. Along with the rapid evolution comes the risks, threats and vulnerabilities in the system for those who plan to exploit it. Therefore, ensuring fool proof end-to-end (E2E) security becomes a vital concern. Artificial intelligence (AI) and machine learning (ML) can play vital role in design, modelling and automation of efficient security protocols against diverse and wide range of threats. AI and ML has already proven their effectiveness in different fields for classification, identification and automation with higher accuracy. As 5G networks' primary selling point has been higher data rates and speed, it will be difficult to tackle wide range of threats from different points using typical/traditional protective measures. Therefore, AI and ML can play central role in protecting highly data-driven softwareized and virtualized network components. This article presents AI and ML driven applications for 5G network security, their implications and possible research directions. Also, an overview of key data collection points in 5G architecture for threat classification and anomaly detection are discussed. 

\end{abstract}

}

\maketitle

\IEEEdisplaynontitleabstractindextext
\IEEEpeerreviewmaketitle
\IEEEpeerreviewmaketitle
\vspace{-0.5cm}
\section{Introduction}
\label{introduction}
\IEEEPARstart{T}{he} continuously evolving communication network architecture to integrate diverse range of devices with unique requirements for different network parameters has resulted in sophisticated challenges for network security. The recent developments in 5G Networks and beyond are facilitating the immersive growth of data communication by providing higher data rates and speeds. Such gigantic increase in data traffic and connected devices means more vulnerabilities, threats, and attacks resulting in catastrophic damages financially, socially and on humanity. Therefore, scrutinizing and analysis of such Big Data for suspicious activities can not only be achieved with traditional/typical methods. In this context, Artificial intelligence (AI) and Machine Learning (ML) \cite{you2019ai,yao2019artificial} are envisioned to play a key role in solving previously considered NP-hard, and complex optimization problems. The Self-Organizing networks, intelligent and adaptive algorithms implemented in different parts of network architecture paved the way for use of AI and ML with even higher performance gains at smaller costs. The ITU has also established a standard Y.3172, which outlines the architectural framework and requirements for different use cases of ML in future networks including IMT 2020.

\begin{figure*}[!htb] \centering
\includegraphics[width=\linewidth, height=10cm]{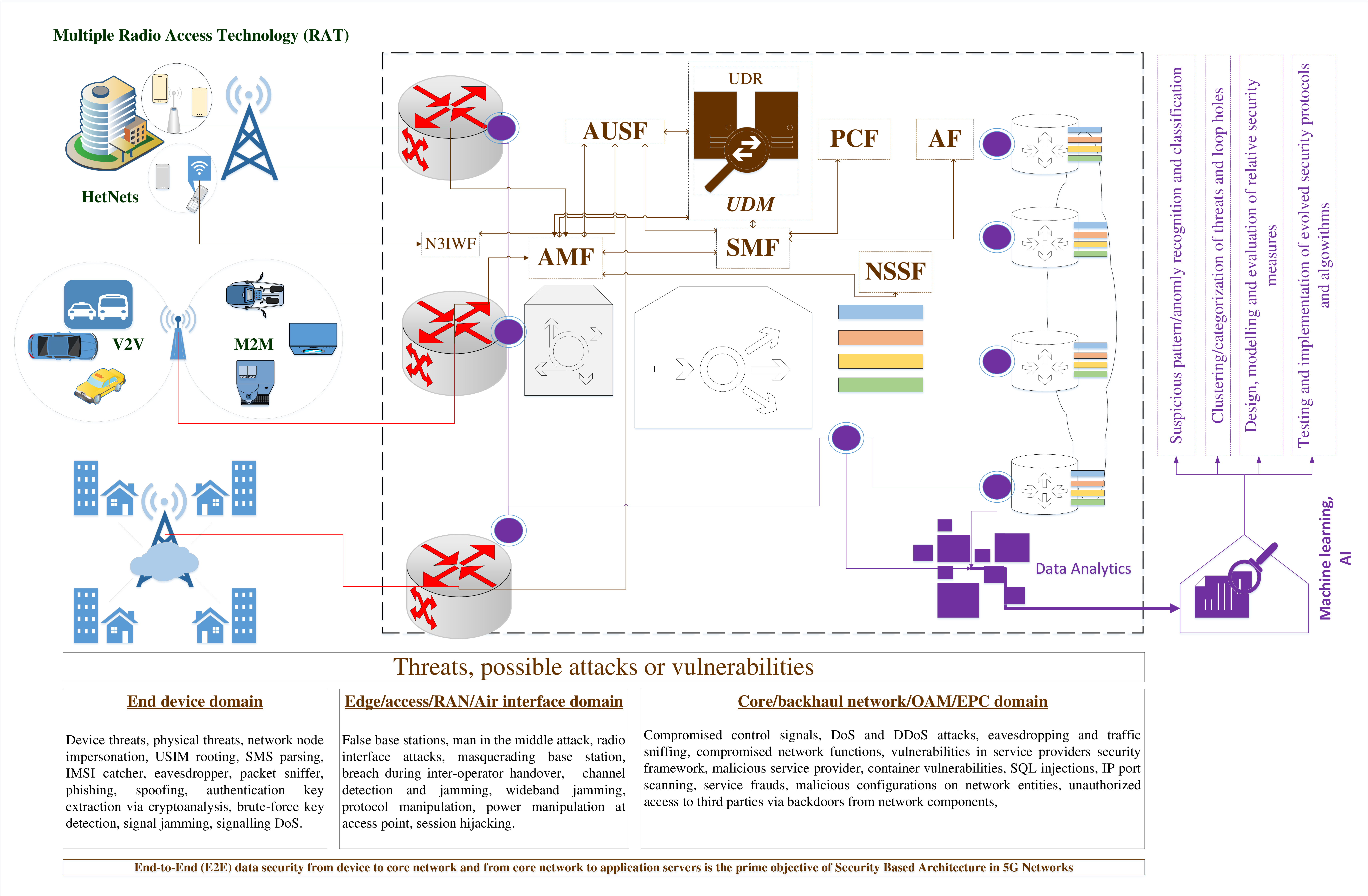}
\caption{Envisioned applications of AI and Machine-Learning in 5G Network Architecture.}
\label{5GArchiAI_ML}
\end{figure*}

Digital bandit have also proven their penetration skills even to most secure and encrypted networks by exploiting vulnerabilities. Such vulnerabilities lead to data theft, cyber attacks, infrastructure damage, ransom demands, blackmailing, disruption of critical services, threats to democracy and fatal to human lives often reported as breaking news. Thus, increasing the necessity to also invest in methods which enables safer and secure communications with transparent user policies, trust models and End-to-End (E2E) visibility. Moreover, 5G and beyond future networks architecture has seen a paradigm shift from the concept of dedicated networks resources for dedicated network functions to more dynamic virtualization, cloudificiation, orchestration, automation and softwareization of network functions from common/shared network resources \cite{khan2019survey}. These factor impose a greater risk to network security and user data if the safety protocols are unable to not only detect threats and attacks but also prevent them in real-time (with minimum delay). Such real-time threat/anomaly detection in terabytes of data require assistance of AI and ML. The data collection points can be set in different parts of the network from access to core network and fed to ML/AI engines for real-time threat detection and attack prevention. Fig. \ref{5GArchiAI_ML} shows the envisioned architecture for integrating AI and ML to detect threats for classification and testing of security protocols against detected threats/attacks in 5G and future networks. Such AI and ML assisted network security can provide cost efficient and sustainable solutions.  

The organization of the upcoming content of the article is as follows. Section \ref{section2} highlights the abstract level details of threats, attacks, and vulnerabilities at different points in 5G networks along with the latest developments and standardization activities related to 5G and future networks security. Section \ref{section3} discusses the taxonomy of AI and ML related technologies along with their implementation gains. Then, Section \ref{section4} presents opportunities, use cases, applications and advantages of different field of AI and ML in 5G security. Section \ref{section5} presents challenges and possible future research directions of AI and ML assisted network security. Finally conclusions are given in Section \ref{section6}. 

\section{5G Networks Security}
\label{section2}

The 3GPP Technical Specifications Group Services \& Systems Aspects (TSG SA3) in its Release 14 highlighted the 17 key threat / areas and possible solutions for security architecture of 5G networks. The security architecture, procedures and requirements for 5G systems were then formulated in Release 15 (R15) in June 2019 \cite{3GPPR15}. The R15 includes security standards for standalone and non-standalone Enhanced Mobile Broadband scenarios, whereas, upcoming R16 and R17 will be focusing on security standards for massive Machine Type Communication and Ultra Reliable Low Latency Communications. The new security features aims to provide E2E security along with flexibility of incorporating multiple authentication frameworks, and higher-layer security protocols to support security for Service Based Architecture (SBA) in 5G. The SBA and network slicing in 5G networks allows higher modularity in the design measure of security protocols. The E2E security architecture can be segregated into two groups. The first one named Network Access Security defines procedure and requirements of securely connecting end-device to radio access network. These procedures secure the device connectivity from end device and edge/RAN domains threats as shown in Fig. \ref{5GArchiAI_ML}. From hereon, ensuring protection of data and privacy from access network to core network and beyond can be referred as Network Domain Security.

\begin{figure*}[!htb] \centering
\includegraphics[width=\linewidth, height=08cm]{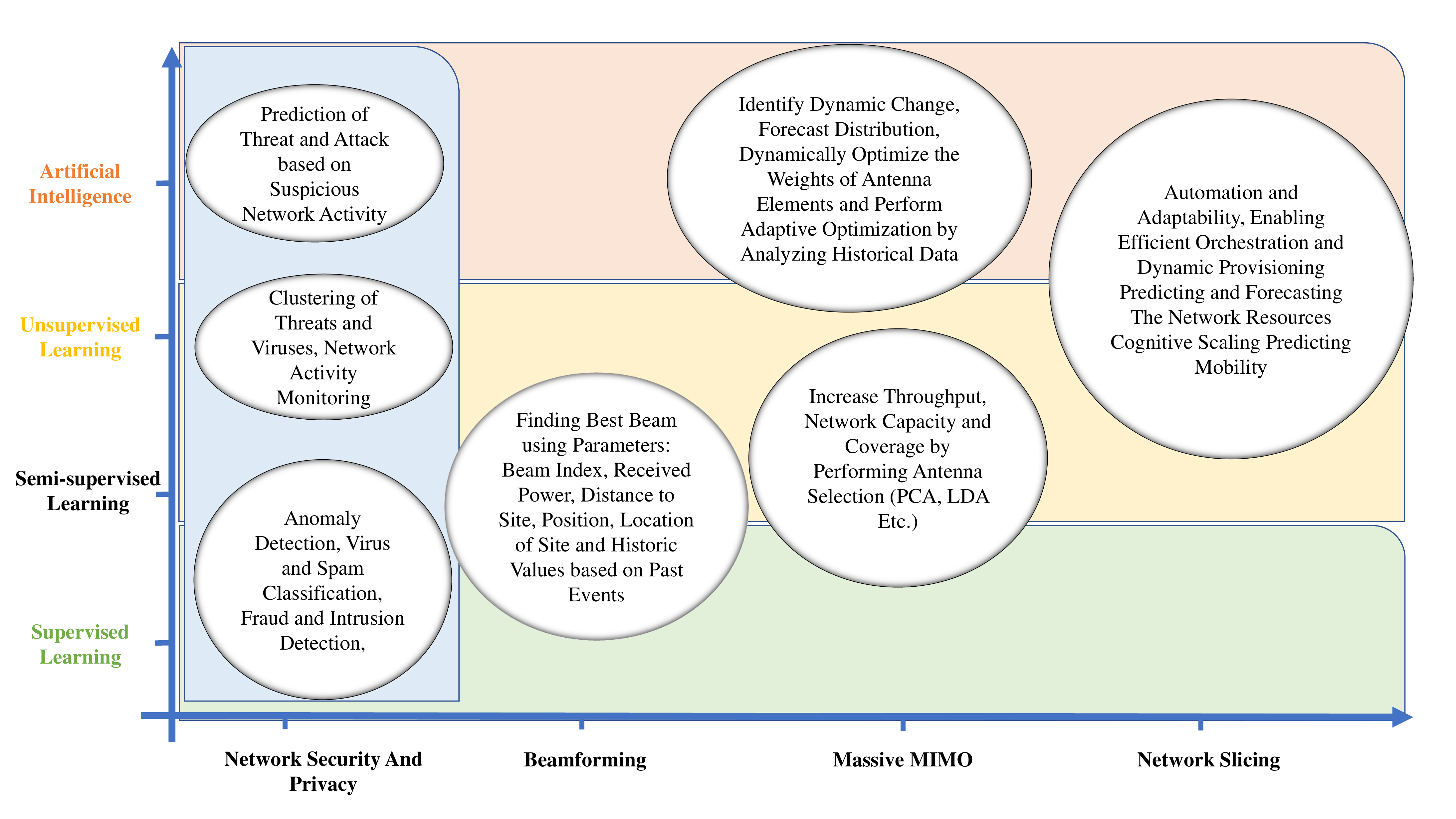}
\caption{Applications of Artificial Intelligence and Machine Learning in 5G network }
\label{AI_ML_application}
\end{figure*}

Highly software-centric and dynamic 5G network architecture where user data is traversing through several network slices and layers, also require agile, adaptive and robust security management and automation. Depending on different network slices for different services, the security requirements are also different from lightweight, middleweight to heavyweight security. These hierarchical security levels suiting needs of different slices can more easily be implemented with software-based evolving techniques. Contrary to this, manual or traditional need based upgrades to network security are no more feasible, therefore, security automation should be an integral part of the overall network. Leading industry partners are now planning to leverage AI and ML to incorporate network security for 5G and beyond wireless networks.

Recent advancements in AI and ML can enhance the performance of next-generation 5G networks. AI and ML have opened gateways to new robust and dynamic solutions in the domains of security, privacy, and threat detection in 5G systems. AI and ML has shown significant potential in terms of performance gains for wireless systems in the domains of beamforming, massive multiple input multiple output (MIMO), and network slicing. Different use cases and possible applications of AI and ML for 5G are also shown in Fig. \ref{AI_ML_application}. 
\section{Artificial Intelligence and Machine Learning}
\label{section3}

The concept of using AI and ML in security and privacy is not new but their feasibility and performance superiority gained attention with  the evolution of deep learning (DL) algorithms. Most of the methods before the development of DL were dedicated to model the attack patterns with certain characteristics that are not robust in nature, but with deep AI and ML, it is expected that systems will become more resilient towards new sophisticated threats and attacks with dynamic characteristics. Because, attackers use sophisticated techniques like obfuscation, polymorphism or impersonation to avoid detection. From packet capturing and analysis to big data insights, AI and ML can be leveraged to notify the threats not detected by conventional techniques. The pattern-based learning at the core supported by softwarization and virtualization provides agility and robustness to timely counter the threats and attacks.

AI is showing a positive impact on the information security field. AI algorithms are being adopted to address security and privacy issues. The information security industry is generating more and more data that opens them to advance threats and AI could be a powerful antidote. The first generation of AI solutions are focusing on scrutinizing data, detect threats and assist humans in the remediation plan. The second generation of AI will make the systems more autonomous and only leave the critical support issues to humans \cite{lee2017security}.

\subsection{Possibilities of AI and ML in 5G}
An increased bandwidth, higher spectrum utilization and high data rates in 5G networks have also widen the threat and privacy landscape from personal device to the service provider network. Thus, the network should be smart enough to deal with these challenges in real-time and ML and AI techniques could help model these robust dynamic algorithms that can help to detect network issues and provide with the possible solution in real-time. In the same way, AI and ML protect the personal devices that are connected to the internet by providing adaptive security solutions that can tackle diverse network situations, threats, and attacks. In short to medium term plan, AI and ML can be used to detect the threats and counter them with the robust and adaptive security algorithms. Whereas, in the long-term, a fully automated security mechanism is envisioned for timely response to threats and attacks.

The 5G networks are expected to support much higher level heterogeneity (in terms of connected devices and networks) as compared to its predecessors. For instance, 5G networks support smart vehicles, smart homes, smart buildings and smart cities. Similarly, the Internet of Things (IoT) in 5G network structure will involve more robust and adaptive techniques to handle the critical security issues both at the network and device sides. The security of such networks will be much more complicated because of the outside intrusion as well as the local intrusion. AI and ML can provide solutions by classifying fragile security links in-between, for instance, identity, authentication, and assurance. The security and privacy in 5G-IoT will cover all the layers such as identity protection, privacy, and E2E protection. For instance, the key authentication framework from end-device to core network and on-ward to service provider, while concealing the key identifier is still a complex issue. We believe AI and ML can also play an important role in key authentication along with effectively minimizing the masquerading attacks.

Catering for security and privacy of data from these different systems with uniquely different security requirements become a tedious task. Powerful AI and ML with overview of SBA and security requirements for different end-systems can can detect and rectify these issues in real-time by classifying and clustering unusual threats. This, in turn, greatly assist the workforce skills shortage in information security industry. AI and ML can help in developing security mechanisms by creating trust models, device security and data assurance to provide systematic security for the whole 5G-IoT network. 

\begin{figure*}[!htb] \centering
\includegraphics[width=\linewidth, height=8cm]{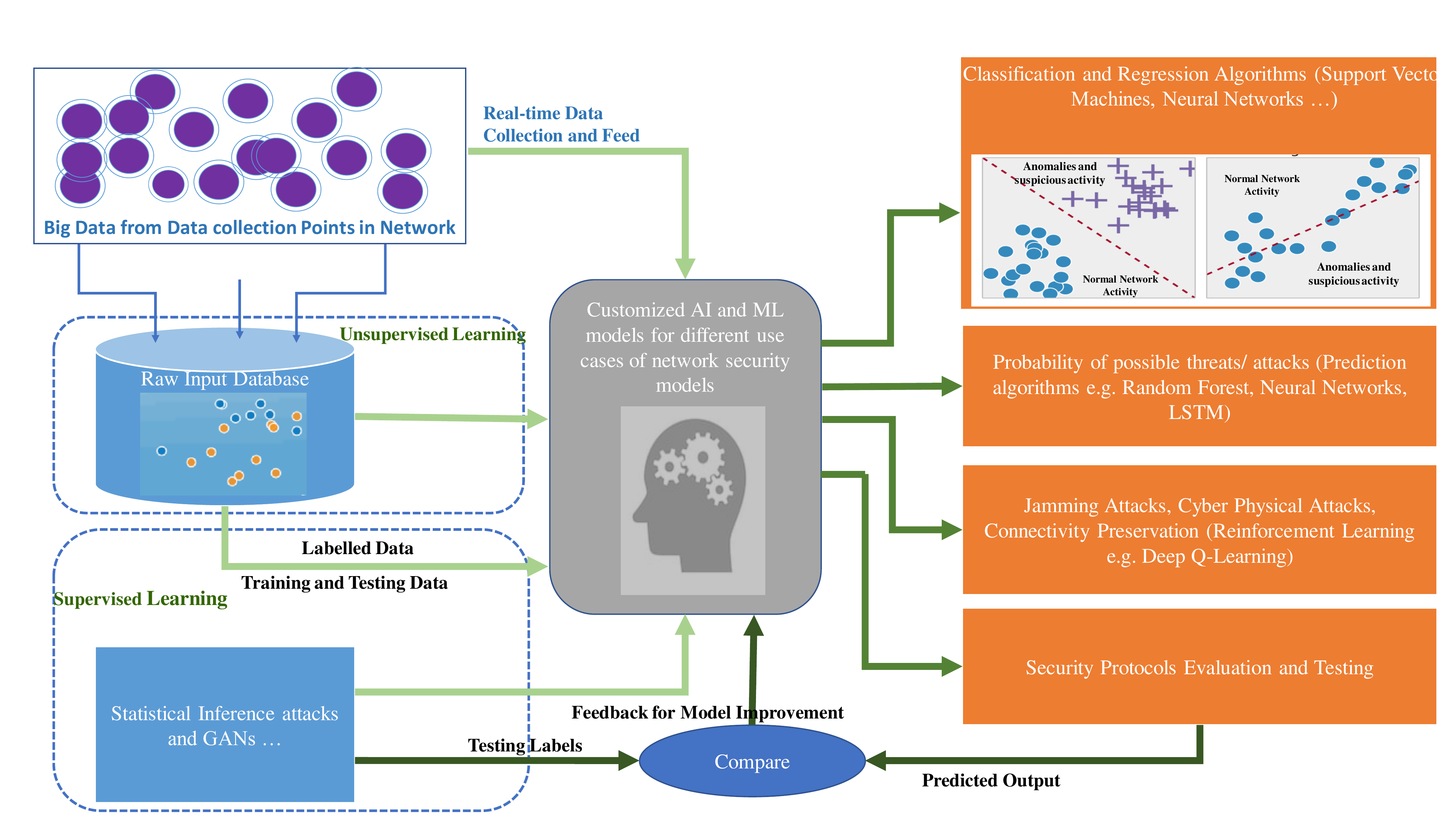}
\caption{Different application scenarios and use cases of AI and ML assisted Network Security.}
\label{AI_ML_usecases}
\end{figure*}

\section{Applications of AI and Machine Learning for 5G Security}
\label{section4}

Mostly, AI and ML algorithms are data-hungry in nature which means that data is needed to train the model for effective functioning. In the era of 5G, data generation, storage, and management is not difficult as we have high computational power, exponential data growth, and data sources. The network can be maintained, accessed and analysed for possible threats, attacks and vulnerabilities using AI and ML at a lower cost of computing, and affordable infrastructure.  Fig. \ref{AI_ML_usecases} summarized the various applications of AI and ML in network security. AI and ML models can be used to detect suspicious activities in real-time by analysing network activity patterns and parameters. Classification algorithms can be used to detect anomalies by monitoring network parameters such as throughput and network error logs. Clustering algorithms can be used to categorize various kinds of threats and loopholes in network security. The models such as statistical inference attacks and generative adversarial networks (GAN) can generate fake datasets to develop and evaluate new security measures as well as testing and implementing evolved security protocols and algorithms.

The research in developing private AI and ML models have seen some significant progress in secure computation, encryption, privacy, and federated learning. Hybrid models are created by adopting techniques from different fields to make models efficient, faster and generalized. The most common and popular example is the differential privacy introduced by Google security and privacy team \cite{hassan2019differential}. The secure computation field is making new progress by aggregating distinct protocols for faster computations \cite{Riazi2018CHS}. Some examples include Gazelle, TAPAS, and Faster CryptoNets that are used for secure computation with homomorphic encryption.  SecureNN is an ML solution that uses comparison-based operation of neural networks for bit extraction and secret sharing \cite{tanuwidjaja2019survey}. Federated learning and secure enclaves are also using AI and ML-based models such as Slalom, Chiron and Ekiden.

Another recent trend is the development of generic and robust anomaly detection algorithms which deals with unknown attacks \cite{bhuyan2013network}. AI and ML can be applied to deal with most of the applications such as antivirus scanner systems, intrusion detection, spam filters, and fraud detection systems. The methods generally work on data generated by network traffic, host processes, etc. Unsupervised algorithms such as neural networks and clustering can be used to assist humans in identifying suspicious activities.

The 5G and beyond network relying on SBA, independent decentralized network functions and third-party servers will pose a greater threat in terms of Denial-of-Service (DoS) and cyber-attacks. Thus, dedicated agents according to the domain of network components can better safe-guard these components in particular and overall system in general. Various AI and ML solutions have been presented to deal with decentralized networks \cite{dai2019blockchain}. The recent solutions are using several reinforcement learning (RL) and deep  reinforcement learning (DRL) techniques to deal with such attacks \cite{luong2019applications}. In case of jamming attacks where, the hackers jam the radio frequency (RF) signals, DRL based solutions were developed that select appropriate frequency channels and avoid attack using an optimal policy learned on previous observations. Cyber-physical attacks manipulate data to gain control of the system. These kinds of attacks usually occur on autonomous systems such as smart vehicles \cite{ferdowsi2018robust}. DRL provides autonomous systems the ability to learn from the time-varying observations to generate optimal actions so that the system can be more robust and dynamic. DRL systems show decent progress in connectivity preservation among robots to support efficient communication .

Deep Learning (DL) is also providing its benefit to cybersecurity solutions as it can automatically learn patterns from past entries to avoid future intrusion and identify irregular patterns. DL has been successfully deployed in infrastructure-level security (anomalies detection on the physical network), software-level security (malware, virus and botnet detection in the mobile network) and user-level security (private information protection) \cite{zhang2019deep}. Different variations of DL networks such as auto-encoders, dense networks, and Convolutional Neural Network are used in several security applications including malware detection, DoS probing, flooding, instant signal-to-noise ratio variations, and various other cyber-attacks. At the software-level, DL networks are used in the classification of malicious applications, spams, unknown traffic, and botnet. In the case of user privacy, DL has shown its potential in data sharing problems, information leakage, and privacy preservation. Table \ref{table1} summarizes the key enabling technologies among AI and ML for potential security applications along with their advantages.

\begin{table*}
\caption{Summary of AI-assisted technologies and the possible case scenarios along with their advantages and challenges}
\label{table1}
\resizebox{\textwidth}{!}{
\begin{tabular}{|l|l|l|l|}
\hline
\rowcolor[HTML]{EFEFEF} 
\textbf{\begin{tabular}[t]{@{}l@{}}AI/ML\\ methodologies\end{tabular}} &
  \textbf{\begin{tabular}[t]{@{}l@{}}Key\\ techniques\end{tabular}} &
  \textbf{\begin{tabular}[t]{@{}l@{}}Key features and applications \\ in network security\end{tabular}} &
  \textbf{Advantages} \\ \hline
\begin{tabular}[t]{@{}l@{}}Supervised\\ learning\end{tabular} &
  \begin{tabular}[t]{@{}l@{}}Bayesian classification.\\ K-Nearest Neighbor (KNN).\\ Neural Networks (NN).\\ Generative Adverserial \\ Network (GAN).\\ Support Vector Machine (SVM).\\ Decision Tree (DT) classification.\\ Recommender System.\end{tabular} &
  \begin{tabular}[t]{@{}l@{}}Classification and regression-based security \\ algorithms design. \\ Identity fraud detection and email spam\\ detection.\\ Risk and threat assessment.\\ Pattern recognition and computational \\ learning theory.\\ Security algorithm design,  development \\ and update.\\ Algorithms for anomaly detection.\\ Packet level analysis for packet-level security\\ framework.\\ Distributed Denial of Service (DDoS) \\ detection and prevention.\end{tabular} &
  \begin{tabular}[t]{@{}l@{}}Software-centric security for\\ heavily software-driven network.\\ Flexible algorithm modelling \\ with evolving functionality.\\ Adaptive security management\\ and automation.\\ Overcoming the workforce and \\ skill shortage with automation.\\ Resolves complex optimization \\ problems.\\ Agile and self-evolving design \\ of security mechanisms.\\ Reduced cost of security operations.\end{tabular} \\ \hline
\begin{tabular}[t]{@{}l@{}}Unsupervised\\ learning\end{tabular} &
  \begin{tabular}[t]{@{}l@{}}Hierarchical clustering.\\ Reinforcement learning.\\ Dimensionality reduction.\\ Association analysis.\\ Hidden Markov analysis.\\ Big data visualization.\end{tabular} &
  \begin{tabular}[t]{@{}l@{}}Malicious content detection from \\ incoming/outgoing traffic analysis.\\ Segregation of legitimate and illegitimate users \\ and traffic.\\ Fully automated grouping/clustering from \\ immensely large traffic data patterns. \\ Security framework optimization from a limited\\ group of data sets (traffic patterns).\\ Application/network slice-based traffic steering.\\ Powerful tools of  analyzing, monitoring and \\ checking on-going traffic.\end{tabular} &
  \begin{tabular}[t]{@{}l@{}}Automated clustering from highly\\ dynamic data sets.\\ Association mining of features \\ based on common traits.\\ Real-time implementation.\\ Discover unusual data points.\end{tabular} \\ \hline
\begin{tabular}[t]{@{}l@{}}Reinforcement \\ learning\end{tabular} &
  \begin{tabular}[t]{@{}l@{}}Real-time decisions.\\ Robot navigation.\\ Q learning.\\ Deep Q learning.\\ Skill acquisition.\\ Game AI.\end{tabular} &
  \begin{tabular}[t]{@{}l@{}}Automated actions  based on the severity of \\ detected events or breaches.\\ Automatic adaptation for updated data patterns.\\ Pattern driven decisions and predictions\\ for future attacks.\end{tabular} &
  \begin{tabular}[t]{@{}l@{}}Highly robust and trained agent \\ for timely decision making.\\ Efficient for mission-critical and \\ delay-sensitive digital infrastructure.\\ Higly adaptable for tackling with\\ diverse set of threats.\end{tabular} \\ \hline
\end{tabular}
}
\end{table*}

\section{Open Research Challenges and Future Directions}
\label{section5}

The aim of fully automated cyber defense system might be a long-term goal, but meanwhile, the cost of integrating AI and ML for existing and future systems also need rigorous analysis. Moreover, it is also being reported that cyber-hackers have also started taken advantage of AI and ML based smart algorithms for attacks and vulnerability exploitation. As, implementation studies are still being conducted on safe, smart and powerful integration of AI and Ml, some fundamental challenges still need a critical research and analysis. For instance, in finding anomalies in a network, first, we need to define normal. Network activity is seldom normal, and therefore, a fully supervised or semi-supervised network would be one possible way to deal with in this situation. ML models use large chunks of data to learn and make pattern for regression on unseen data. If the network parameters are changed drastically or variations are to be found, the network will collapse during deployment and therefore, a retrain of the network will be required. Most of the ML methods including DL are black box in its hidden layers and therefore, the insight of its formulation on the trained data is limited in nature. Data privacy is one of the most important issues as AI and ML algorithms feed on data. The use of data in ML increases the risk for an attack as models are trained on the data which can be used for data mining purposes as well. 

ML-based security solutions are always vulnerable to new types of sophisticated attacks such as GANs. Researchers have tested the vulnerability of ML-based security models using simulated GANs \cite{huang2011adversarial}. 

5G-IoT security and privacy needs more investigation in the domains of authentication, authorization, access control, and privacy-preserving. The current 3GPP defined networks use functional node specification and abstract interfaces but in 5G IoT, the network itself will serve as core infrastructure and security assurance will be the key challenge to deal with. At this stage, semi-supervised AI-assisted solutions better suits the distributed systems. With the evolution of AI algorithms, these system will become fully automated in the future. 

Another research trend is dealing with eavesdropping in trusted communication over 5G networks. AI and ML could be used in maintaining device security as well as high layer security in IoT. AI and ML models are flexible and scalable security solutions that can consider multiple network layers for trust modeling and identity management, security assessment, and privacy protection as well as energy-efficient.

Recently, GANs are shown to mimic the exact output of a network whilst having no access to the training data. By using generator and discriminator DNNs in a single training mechanism, the networks compete with one another where the generator generates new data samples whereas the discriminator distinguish them as real or fake; settling onto a game theory approach. The final output is a network that can no longer distinguish between the real or fake samples of data and therefore, it can successfully generate new samples of unseen data. This essentially means that a GAN can improvise user authentication mechanism, generate phishing data, flood core network with spam signaling, all without being exposed to the actual network. GAN pose a series of threats to the ongoing development of AI and ML for network security since it can deceive the core network with accurate authentication. 
\section{Conclusion}
\label{section6}
This paper presents AI-assisted technologies, scenarios and application for security of 5G and beyond wireless networks. The highly dynamic traffic patterns, service-based network architecture, distributed network functions and authentication over multiple servers in 5G and beyond networks require relatively robust, agile and fully automated security framework. Such framework is built-upon smart AI technologies. AI can significantly improve the security for distributed ad-hoc setup of network infrastructure providing different network functions. At this stage, semi-automated security framework is more suitable, however, with continuing evolution in AI technologies and feasibility studies of safe implementation of these technologies will decide the end goal of complete automation. Substantial research is needed to address the challenges and issues before AI fully takes over the digital automation. 
\vspace{1.0cm}

\bibliographystyle{ieeetr}
\bibliography{references}

\begin{IEEEbiography}[{\includegraphics[width=1in,height=1.25in,clip,keepaspectratio]{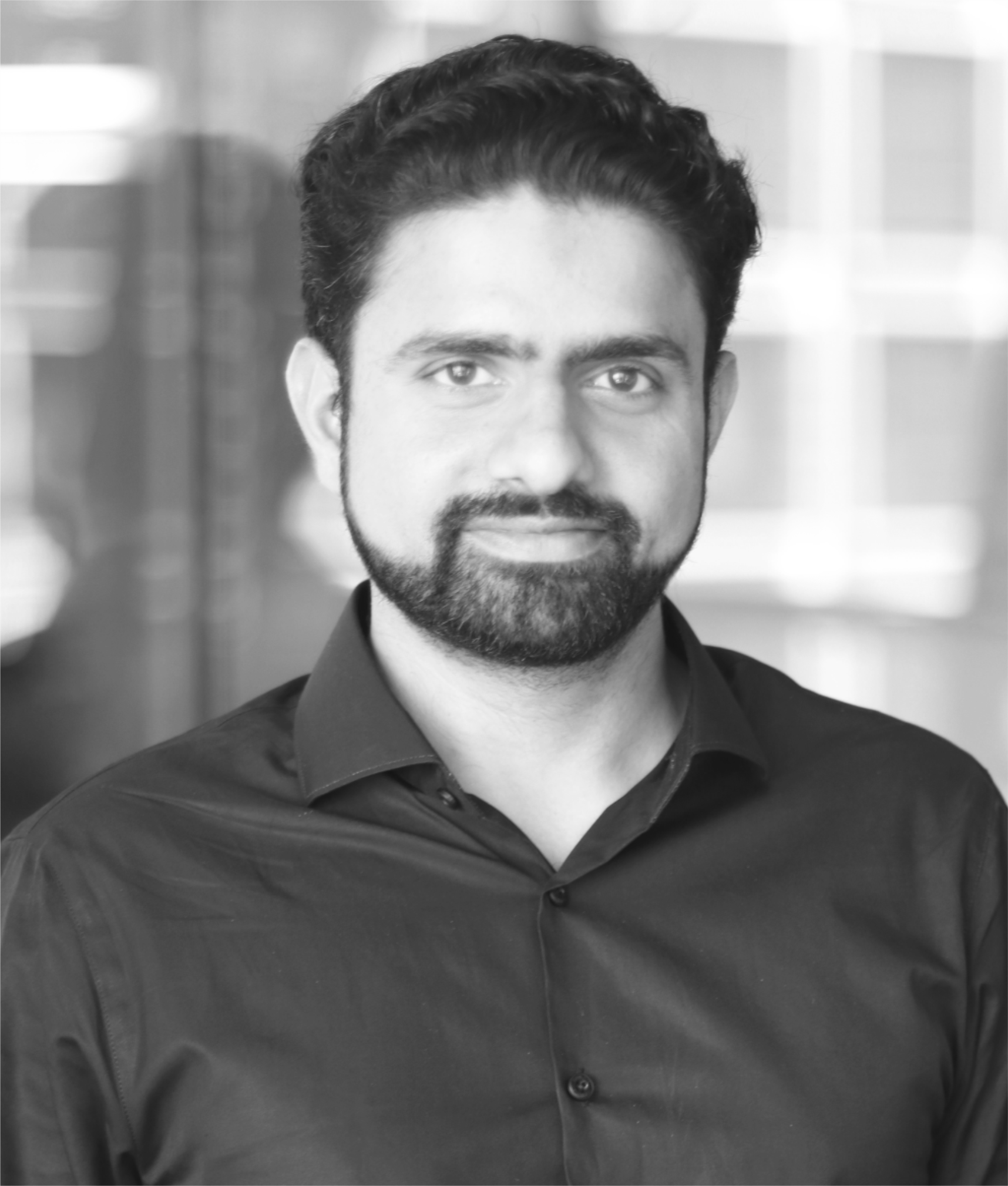}}]{Noman Haider} Noman Haider is currently working as Lecturer at Victoria University, Sydney, Australia. Noman completed Ph.D. and M.Sc in Engineering and Information Technology from University of Technology Sydney, Australia and Universiti Teknologi Petronas, Malaysia in 2019 and 2014, respectively. Noman received B.S. (Electronics Engineering) degree from Mohammad Ali Jinnah University, Pakistan in 2011. From 2012 to 2019, Noman has worked on multidisciplinary research projects in collaboration with professionals from academia and industry (Intel US and Intel Europe). His research interest includes resource sharing and allocation for future wireless networks, network security, artificial intelligence, and context-aware learning models.
\end{IEEEbiography}

\begin{IEEEbiography}[{\includegraphics[width=1in,height=1.25in,clip,keepaspectratio]{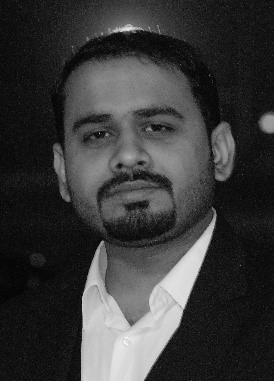}}]{Muhammad Zeeshan Baig} Muhammad Zeeshan Baig is a Ph.D from the Department of Computing, Macquarie University, Australia. He also worked as a visiting research scholar at Northumbria University, UK. He has received a scholarship from a European Unions Erasmus Mundus external cooperation programme called cLINK (Centre of Excellence for Learning, Innovation, Networking and Knowledge). His research interest includes Artificial Intelligence, Machine learning biomedical image and signal processing, brain-computer interface, and machine learning.  
\end{IEEEbiography}

\begin{IEEEbiography}[{\includegraphics[width=1in,height=1.25in,clip,keepaspectratio]{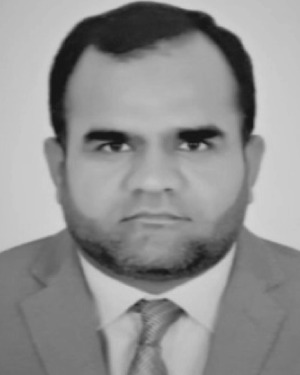}}]{Muhammad Imran} Muhammad Imran is working as an Associate Professor in the College of Applied Computer Science, King Saud University. His research interest includes mobile and wireless networks, Internet of Things, cloud and edge computing, and information security. He has published more than 200 research articles in reputable international conferences and journals. His research is supported by several grants. He serves as an associate editor for many top-ranked international journals. He has received various awards.
\end{IEEEbiography}

\end{document}